# Continuous Ultraviolet to Blue-Green Astrocomb

Yuk Shan Cheng,[1] Kamalesh Dadi,[1] Toby Mitchell,[1] Samantha Thompson,[2] Nikolai Piskunov,[3] Lewis D. Wright,[4] Corin B. E. Gawith,[4,5] Richard A. McCracken[1] and Derryck T. Reid[1,*]

[1]Institute of Photonics and Quantum Sciences, Heriot-Watt University, Edinburgh EH14 4AS, United Kingdom; [2]Astrophysics Group, Cavendish Laboratory, J.J. Thomson Avenue, Cambridge, CB3 0HE, United Kingdom; [3]Department of Physics and Astronomy, Uppsala University, Box 516, 751 20 Uppsala, Sweden; [4]Covesion Ltd, Unit F3, Adanac North, Adanac Drive, Nursling, Southampton, SO16 0BT, United Kingdom; [5]Optoelectronics Research Centre, University of Southampton, Southampton, Hampshire SO17 1BJ, United Kingdom.

*D.T.Reid@hw.ac.uk

## Abstract

The characterization of Earth-like exoplanets and precision tests of cosmological models using next-generation telescopes such as the ELT will demand precise calibration of astrophysical spectrographs in the visible region, where stellar absorption lines are most abundant. Astrocombs—lasers providing a broadband sequence of ultra-narrow, drift-free, regularly spaced optical frequencies on a multi-GHz grid—promise an atomically-traceable, versatile calibration scale, but their realization is challenging because of the need for ultra-broadband frequency conversion of mode-locked infrared lasers into the blue-green region. Here, we introduce a new concept achieving a broad, continuous spectrum by combining second-harmonic generation and sum-frequency-mixing in an aperiodically-poled MgO:PPLN waveguide to generate gap-free 390–520 nm light from a 1 GHz Ti:sapphire laser frequency comb. We lock a low-dispersion Fabry-Pérot etalon to extract a sub-comb bandwidth from 392–472 nm with a spacing of 30 GHz, visualizing the thousands of resulting comb modes on a high resolution cross-dispersion spectrograph. Complementary experimental data and simulations demonstrate the effectiveness of the approach for eliminating the spectral gaps present in second-harmonic-only conversion, in which weaker fundamental frequencies are suppressed by the quadratic $\chi^{(2)}$ nonlinearity. Requiring only $\approx$100 pJ pulse energies, our concept establishes a practical new route to broadband UV-visible generation at GHz repetition rates.

## 1. Introduction

Solar-like stars exhibit the highest density of stellar absorption lines in the blue-green spectral region, making this part of the spectrum particularly rich in radial velocity (RV) information, where each line is used to help calculate the Doppler shift of the star, and the use of more lines equates to an improved RV error [1]. Improvements to RV measurements become particularly important for the detection of very small RV signals on a long-period baseline, such as the 10 cm s$^{-1}$ amplitude sensitivity required in the search for Earth-like exoplanets, and for precision cosmological observations like the Sandage-Loeb Test [2]. Astrocombs—broadband laser frequency combs with multi-GHz spacing—represent the ultimate wavelength calibration source for this task, by virtue of their stability, atomically-referenceable accuracy, sub-100-kHz linewidths, multi-GHz spacing and potential for broadband spectral coverage [3,4].

Blue-visible astrocombs have previously been demonstrated using a variety of approaches. Second harmonic generation of near-infrared mode-locked lasers has produced narrow-band astrocombs at 400 nm [5,6]. Supercontinuum generation in photonic-crystal fiber has enabled blue-green (435–600 nm, [7]), green (530–560 nm, [8]), and green-red (500–620 nm, [9]) astrocombs. Recent improvements in the manufacture of silicon nitride (SiN) and periodically-poled lithium niobate (PPLN) waveguides have led to on-chip approaches which exploit $\chi^{(2)}$ and $\chi^{(3)}$ processes. Obrzud *et al.* [10] demonstrated 400–600 nm of gap-free coverage at 10 GHz mode spacing through $\chi^{(3)}$ triple-sum-frequency-generation of an electro-optic modulator comb in SiN, however the intrinsically wide mode spacing of the pump source required amplification, broadening and compression stages to achieve sufficient peak power for nonlinear conversion, despite the enhancement offered by waveguide confinement. Nakamura *et al.* [11] reported a 30 GHz astrocomb which extended down to 350 nm via second, third and fourth harmonic generation of a 1.5 μm source in a chirped PPLN waveguide, however wide gaps remained in the spectral coverage and the 250 MHz pump laser required multiple filtering and amplification stages to achieve the desired mode spacing. A similar approach has recently been demonstrated using an EOM pump source [12].

In this work, we propose and experimentally demonstrate a new approach that achieves a broad, continuous visible spectrum by combining second-harmonic generation (SHG) and sum-frequency-mixing (SFM) in an aperiodically-poled MgO:PPLN waveguide. On its own, SHG suppresses weaker spectral features contained in the fundamental light, since it is a quadratic ($\chi^{(2)}$) nonlinearity, but by using a strong auxiliary pulse, SFM can be used to linearly transfer weak but broad infrared components into the visible. We illustrate this experimentally by using a Ti:sapphire laser frequency comb operating with a repetition frequency of $f_{rep} = 1$ GHz to generate 390–520 nm light from an infrared supercontinuum, achieving a gap-free frequency bandwidth of >190 THz. The wide mode spacing of this visible comb means that a relatively low finesse Fabry-Pérot cavity can then be used to extract a sub-comb spaced at a high multiple of $f_{rep}$, and providing strong suppression of nearest-neighbor 1-GHz modes. We implement such a scheme by locking a low-dispersion Fabry-Pérot filter cavity to the incident comb modes to produce a 30 GHz astrocomb spanning 392–472 nm and limited only by the mirror reflectivity bandwidth. The resulting comb is visualized on a comb-mode-resolving cross-dispersion echelle-prism spectrograph, demonstrating well resolved comb lines across all 24 diffraction orders.

## 2. Visible astrocomb generation concept

While $\chi^{(3)}$-mediated supercontinua are readily available in the infrared, with few-pJ octave-spanning generation being reported [13], equivalent performance in the ultraviolet to blue-green region is more challenging because of the tightly-confined waveguide structures required to achieve the necessary group-delay dispersion (GDD) profile for efficient, broadband generation. Instead, our approach uses a pump pulse to generate a broad, near-infrared supercontinuum and employs SHG and SFM between the supercontinuum and a replica pump pulse to create a broad, gap-free visible comb. Since the supercontinuum light has the same carrier-envelope offset frequency ($f_{CEO}$) as the pump, the SHG and SFM components share identical comb offsets of $2f_{CEO}$ and are indistinguishable within the comb. As Fig. 1a illustrates, when the pump and supercontinuum fields are incident independently on a $\chi^{(2)}$ medium they are converted in a way that suppresses weaker supercontinuum components because of the quadratic SHG process. By temporally overlapping the pump and supercontinuum fields, SFM can be used to bridge such gaps, efficiently transferring the supercontinuum bandwidth into the visible region. However, this process alone is not enough to enable a practical implementation, which also requires an efficient means of realizing $\chi^{(2)}$ conversion across an exceptionally broad bandwidth. For this purpose we use an MgO:PPLN ridge waveguide (Fig. 1b) [14], which offers high coupling efficiencies, strong confinement and broadband conversion in a suitably designed quasi-phasematched grating structure (see Methods). Figure 1d shows the spectra of the modelled fundamental and supercontinuum fields (red shading) and the spectrum of the SHG / SFM field produced by the optimized grating design (blue shading). The delay-resolved spectra shown in Fig. 1e illustrate how the addition of SFM between the pump and supercontinuum fields generates around 100 THz of additional bandwidth having high spectral intensity in the 420–470 nm region.

## 3. Experimental astrocomb preparation and characterization

We employed a Ti:sapphire laser comb delivering 33 fs, 1 GHz pulses at 800 nm (Fig. 2a) to pump a 35-mm-long photonic-crystal fiber (PCF) [15], whose closely spaced zero-dispersion wavelengths of 775 nm and 945 nm and high nonlinearity ($\gamma \approx 130$ W$^{-1}$ km$^{-1}$) produced a smooth 650–1050 nm spectrum (Fig. 2d). After an appropriate optical delay, replica pump pulses were co-launched with the supercontinuum pulses into an MgO:PPLN waveguide (Fig. 2a, inset). The waveguide was driven by pump and supercontinuum pulses with energies of 300 pJ and 100 pJ respectively (Fig. 2d, red shading), and produced several milliwatts of ultraviolet to blue-green output from 390–520 nm (Fig. 2d, blue shading). These spectra broadly match the original design predictions (Fig. 1d), but in practice a weaker infrared supercontinuum component reduced the blue generation above 450 nm. Figure 2b shows the experimentally measured delay-resolved SHG / SFM spectrum, which was qualitatively similar to the *a priori* design in Fig. 1e. The ultraviolet to blue-green spectrum directly from the waveguide was easily observable visually, as shown in the photograph of Fig. 2c, where the waveguide output was dispersed and imaged on a screen.

An astrocomb with a 30 GHz mode spacing was prepared by coupling the output of the MgO:PPLN waveguide into a Fabry-Pérot etalon composed of two plane-parallel mirrors with ≈99% reflectivity coatings from 390–470 nm, which exhibited very low GDD over the same wavelength range and zero GDD at 430 nm. The 1 GHz mode spacing of the master comb allows good nearest-neighbor comb-mode suppression, while relaxing the requirements on the etalon finesse (≈ 300), in turn enabling the design of mirror coatings that exhibit a low GDD across a wide bandwidth. The etalon was dither-locked (see Methods) to selectively transmit one sub-comb with a spacing of 30 GHz.

An echelle-prism cross-dispersion spectrograph with an estimated resolving power of ≈46,000 (see Methods) was configured to visualize the resulting comb modes. This measurement, presented in Fig. 3, provides direct confirmation that the combination of SHG and SFM is effective in delivering a gap-free field of high-density comb modes from 392–472 nm. Comb modes are present throughout each echelle order, as illustrated by the insets contained in Fig. 3, which show the comb-mode details in magnified regions of the echellogram selected to represent the full spectral range. Furthermore, the astrocomb coverage from 392–472 nm entirely fills the operational range of our Fabry-Pérot etalon (see Methods).

## 4. Discussion

In this work we have introduced a new concept for achieving a gap-free, broadband ultraviolet to blue-green astrocomb, which until now has remained only partially resolved by other technical approaches, which have been unable to offer the necessary combination of mode spacing, bandwidth and gap-free spectral structure. We have demonstrated an astrocomb spanning 392–472 nm with 30 GHz mode spacing, corresponding to over 4000 individually resolvable comb modes across a frequency bandwidth of 130 THz. The mode format is well matched to the calibration requirements of future high resolution astronomical spectrographs, such as HARPS3 [16] and ANDES [17], which require a comb-mode spacing of 32 GHz for wavelengths in the 370–600 nm UB band [18], corresponding, at the shortest wavelength, to one comb line per three resolution elements for a nominal resolving power of $R$>75,000. Our approach is readily extended to provide complete coverage of the important visible band by combining it with proven Ti:sapphire-based schemes utilizing photonic-crystal fiber to produce visible combs covering 450–900 nm [19,20]. More generally, the demonstration presented here is an important step towards obtaining a gap-free astrocomb spanning the complete ultraviolet to near-infrared range, for example the 400–1800 nm range specified in the European Southern Observatory's technical specification for the ELT-ANDES spectrograph [21]. Ti:sapphire lasers operating at 1 GHz are readily extended to this near-infrared range by using optical parametric oscillators [22], and a broadband near-infrared astrocomb has been reported by spectrally broadening such an OPO source in a highly nonlinear fiber [23].

Our results also extend the understanding of MgO:PPLN ridge-waveguide devices for ultraviolet / blue generation [24]. While mechanically fabricated MgO:PPLN waveguides have been shown to be damage resilient and capable of multi-watt average power handling [25], in comparison to emerging thin-film lithium niobate waveguides [26], their larger cross-sectional areas are not intrinsically single-mode for visible wavelengths. Despite this, we have consistently observed a high-quality transverse visible mode in the visible, which has enabled successful coupling into the high finesse Fabry-Pérot etalon used for longitudinal mode selection. We attribute this primarily to the efficient transfer, through the nonlinear conversion process, of the fundamental TE0 pump mode profile into the SHG and SFM fields.

Further opportunities exist for developing the approach we describe here to better match it to the specific requirements of astronomical spectrographs. The MgO:PPLN grating used here was designed from *a priori* modelling (see Methods), however the design could be refined by using experimental pulse measurement data describing the intensity and phase of the fundamental fields. In combination with more sophisticated domain-level optimization strategies such as simulated annealing [27,28], this could enable broader, more intense and flatter spectra to be obtained, in turn demanding less intensive spectral shaping [29] before the spectrograph.

**Methods**

**Quasi-phasematched grating design**

The design of the quasi-phasematched MgO:PPLN grating proceeded by combining an optimization algorithm with a forward model of the nonlinear conversion process [27,28], which we extended to broad bandwidths using a nonlinear envelope description [30, 31, 32]. The grating was designed entirely *a priori*, using a supercontinuum field first computed using a generalized nonlinear Schrödinger equation model [33], whose critical parameters (pulse peak power; fiber nonlinearity) were pre-calibrated experimentally. To limit the search space, the local grating period (Fig. 1c) was parameterized as $\Lambda = \Lambda_1 + (\Lambda_2 - \Lambda_1)(z/l)^{\alpha}$, where the initial and final grating periods are $\Lambda_1$ and $\Lambda_2$ respectively, $z$ is the propagation distance along a waveguide of total length $l$, and $\alpha$ describes deviations from linear chirp. Various optimization strategies are possible, but to maximize the gap-free visible bandwidth we defined a merit function proportional to the total wavelength bandwidth exceeding a minimum threshold in spectral intensity. Simple phase-matching calculations using the refractive-index equations for MgO:PPLN [34] were used to determine initial choices for $\Lambda_1$ and $\Lambda_2$ that bracketed the values corresponding to SHG and SFM of the participating fundamental wavelengths. The grating chirp parameter, $\alpha$, and the pump-supercontinuum inter-pulse delay, $\tau$, were initiated at values which were found manually to yield some SHG and SFM over the desired bandwidth. Optimization was performed in MATLAB® using the Nelder-Mead algorithm. The resulting grating design is shown in Fig. 1c, where light enters the end of the waveguide containing the longest domain periods.

**Fabry-Pérot etalon filtering and locking**

The Fabry-Pérot etalon was a plane-plane configuration with a spacing of ≈5 mm, adjusted to precisely match $c/2mf_{rep}$, where $m = 30$ corresponds to an astrocomb spacing of 30 GHz. The mirror performance is shown in Fig. 4 and the coating design (LaserOptik GmbH) was optimized to achieve a very low group-delay dispersion (GDD) from 390–470 nm without the need to employ a pair of complementary-dispersion coatings. This design achieves an etalon finesse >300 from 393–475 nm (Fig. 4d) and nearest-neighbor mode suppression (Fig. 4e) of nearly - 30 dB over the same range. The UV to blue-green light leaving the MgO:PPLN waveguide was prepared in a collimated beam with a diameter of 5 mm, which slightly overfilled the clear aperture of the etalon. We spectrally conditioned the light to match the etalon's operating wavelength range by reflecting it multiple times on replica etalon mirrors before coupling it into the etalon. By doing this we ensured that the intensity envelope of the filtered spectrum was identical to that of the input (see Fig. 2d), and that all spectral components of the incident comb were strongly filtered.

To enable active stabilization, one of the Fabry-Pérot mirrors was mounted on a fast, short-displacement piezoelectric transducer (PZT) and the other on a slower, longer-displacement PZT. To selectively transmit one family of sub-comb modes spaced at $30f_{rep}$ (30 GHz), the étalon mirror spacing was first set to $\approx 5$ mm, then a few-nm, 10 kHz dither signal was applied to the fast PZT while the slower PZT was scanned until transmission peaks were obtained on a silicon avalanche photodiode (Fig. 2a). The detected signal was demodulated at the dither frequency, low-pass filtered and conditioned by a servo amplifier to stabilize the etalon in resonance with the selected sub-comb. Since the Ti:sapphire SHG (400 nm) component of the UV to blue-green spectrum was strongest, this light was responsible for the detected signal, which conveniently allowed the etalon locking to be pre-configured without the supercontinuum input to the waveguide. The identical comb offset frequencies of the SHG and SFM light mean that when the etalon is configured to mode-filter the pump SHG light, the supercontinuum SHG and pump-supercontinuum SFM outputs are also automatically transmitted. The stabilization technique was robust, capable of remaining in lock for several hours, and benefited from not requiring an auxiliary single-frequency UV / blue laser pre-locked to a comb mode.

**MgO:PPLN waveguide fabrication and operation**

The MgO:PPLN waveguide was manufactured by Covesion using an electric-field-poling process in 0.5-mm-thick MgO:LiNbO₃ to create 0.3-mm-wide aperiodic gratings, followed by process steps of zinc indiffusion and ductile dicing to form ridge waveguides as described in [14]. A device length of 5 mm was chosen to support efficient operation, and ridge widths of 11–13 μm were included to enable optimization of the modal overlap and conversion efficiency. The waveguide facets were anti-reflection coated at 380–520 nm and 800–1200 nm to minimize loss and etalon effects. Pump and supercontinuum light were co-launched into the waveguide using a plano-convex lens, achieving a mode field diameter of approximately 5 μm.

**Cross-dispersion echelle spectrograph**

The spectrograph comprised a 31.6 lines mm⁻¹ echelle grating (Thorlabs GE2550-0363) which was used in a quasi-Littrow configuration [35] with $\alpha \approx \beta \approx 63°$ to maximize the diffraction efficiency across each order. The echelle grating was tilted to achieve a small out-of-plane angle [35] of $\gamma \approx 1.5°$, allowing the vertically-dispersed diffraction orders to be intercepted by an F2-glass prism used close to minimum deviation. The prism provided cross-dispersion in the horizontal plane. following which a concave aluminum mirror with a focal length of 200 mm was used to image the echellogram onto a 20.2 megapixel 8-bit CMOS camera. The CCD sensors used in astronomical spectrographs have much greater dynamic range (typically at least 16 bits) and lower dark noise than the sensor available to us. On such a spectrograph, the dynamic range of our astrocomb would be, in principle, fully resolvable, however strong saturation by the pump SHG light was avoided by reducing the camera gain. The associated echellogram orders (139–143) were extracted from a set of low-gain camera frames before being integrated into the main image shown in Fig. 3.

The resolving power ($R = \lambda/\Delta\lambda$) of a diffraction-grating spectrograph is $mN$, where $m$ is the diffraction order and $N$ is the number of illuminated grating elements. At $\alpha \approx 63°$, the 5 mm incident beam diameter illuminates 348 grating elements, corresponding to $R \approx 46{,}000$ at $\lambda = 420$ nm ($m = 134$). Expressed in frequency, this results in a spectral resolution of 15 GHz, sufficient to resolve individual comb modes at 30 GHz spacing.

**Author contributions**

R.A.M, S.T., N.P. and D.T.R. conceived and designed the experiments; K.D., Y.S.C. and T.M. performed the experiments and prepared the results; L.D.W. and C.B.E.G. designed and fabricated the MgO:PPLN waveguide. All the authors contributed to the writing of the paper.

**Acknowledgements**

The authors gratefully acknowledge grant funding from: the UK Science and Technologies Facilities Council (ST/S001328/1, ST/V000403/1, ST/X002306/1 and ST/N002997/1, ST/V000918/1); the Royal Academy of Engineering (RCSRF1718639 and RCSRF2223-1678); and the Knut and Alice Wallenberg Scholarship 2021.

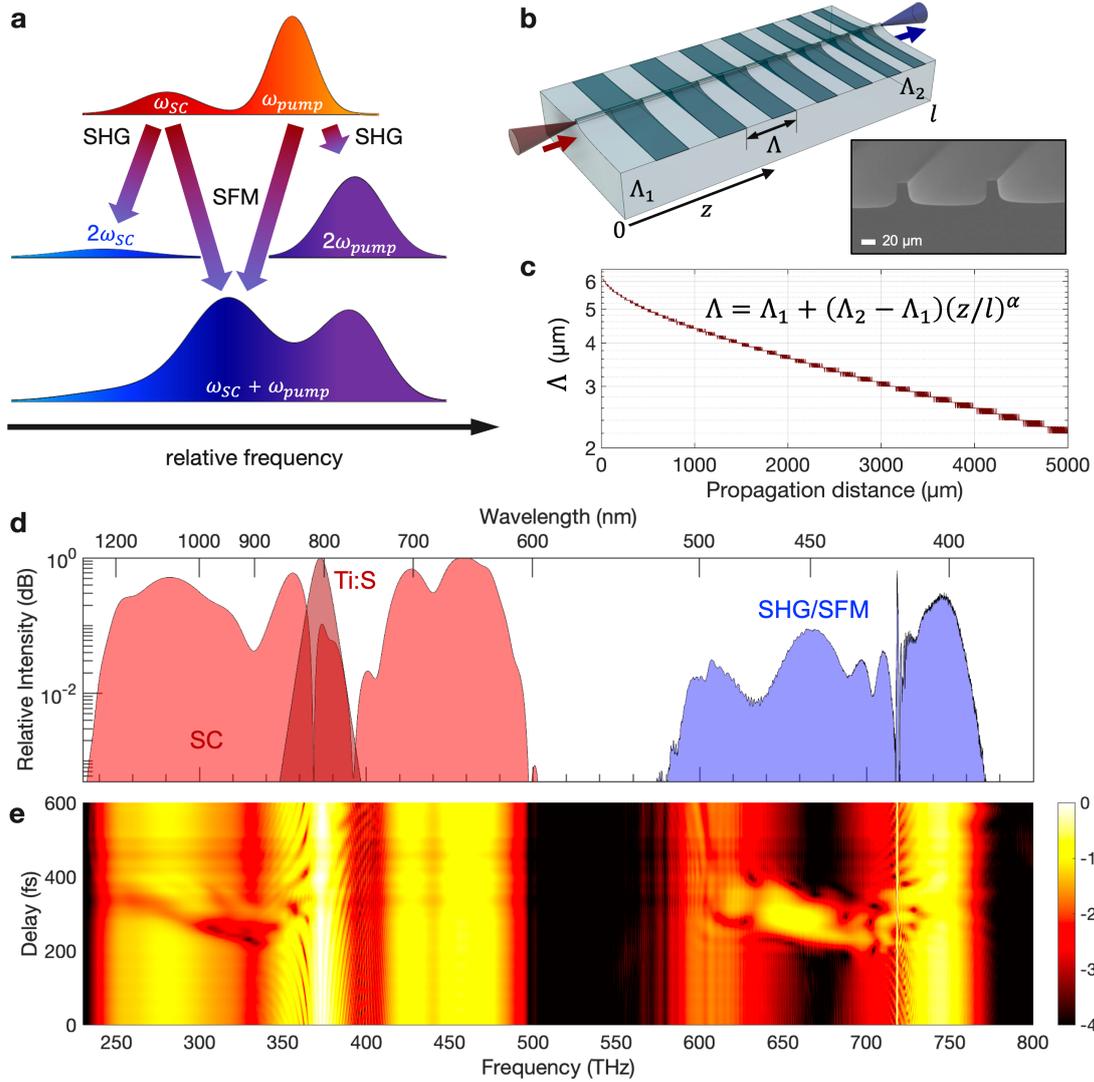

**Fig. 1. Concept of broadband UV to blue-green generation using sum-frequency mixing.**
**a,** The infrared component of a broadband supercontinuum (SC) pulse is mixed in a $\chi^{(2)}$ medium with a higher frequency, stronger pump pulse. In the absence of temporal overlap, only second-harmonic generation (SHG) occurs, with the quadratic nonlinearity suppressing weaker spectral components, and leading to gaps in the generated visible spectrum. By introducing an appropriate delay, allowing the pump and SC pulses to walk through each other, sum-frequency-mixing occurs which blue-shifts and enhances the supercontinuum components in a way that provides gap-free UV-blue generation. **b,** Efficient conversion of 1 GHz repetition-rate, $\approx100$ pJ pulses is achieved by using an MgO:PPLN ridge waveguide containing a custom-designed quasi-phasematched grating. **Inset:** Scanning electron-beam micrograph of a representative waveguide. **c,** The MgO:PPLN domain-width pattern follows an aperiodic design optimized to ensure gap-free upconversion. **d,** Simulated 30-fs pump pulse and supercontinuum (red) and resulting SHG and SFM spectrum achieved for optimal pump-supercontinuum delay ($\tau = 250$ fs). **e,** Simulated delay-resolved spectra illustrating the additional bandwidth created by the SFM process.

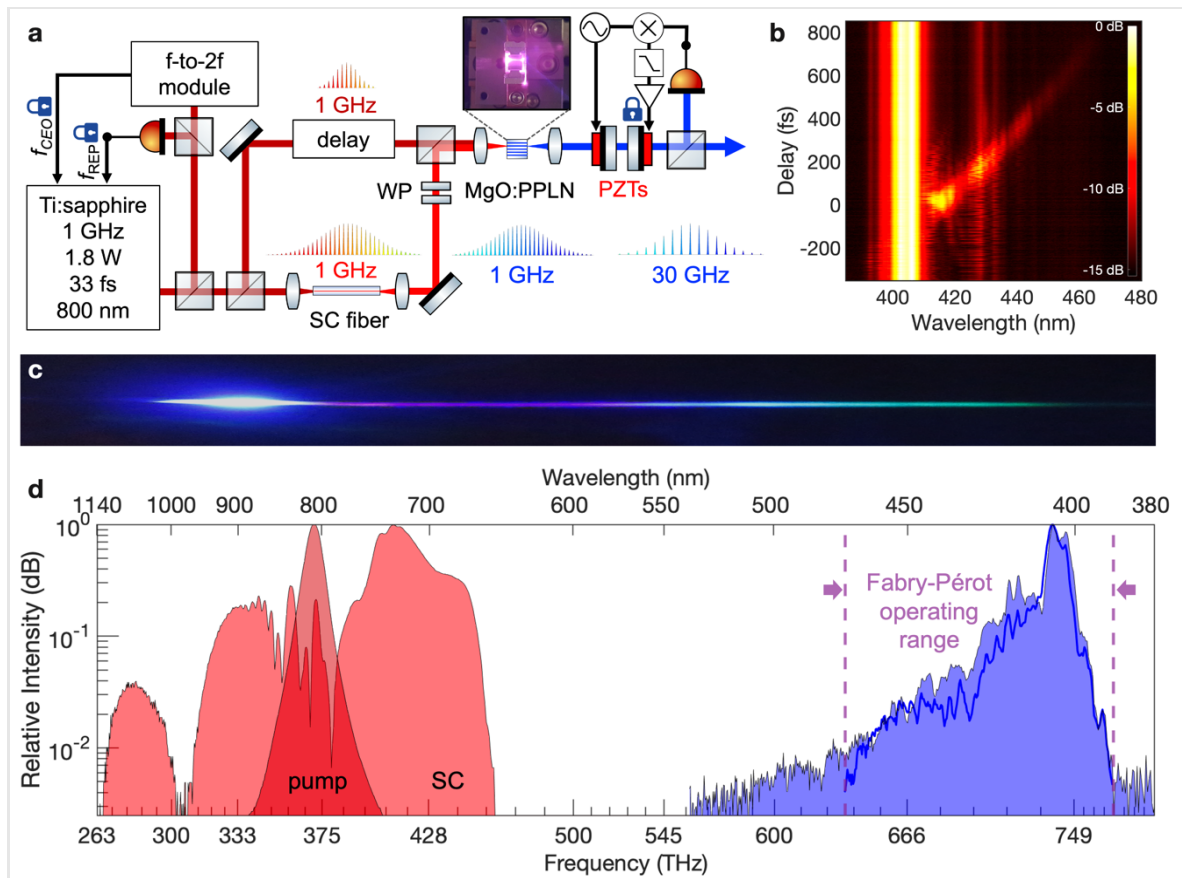

**Fig. 2. Implementation of broadband UV to blue-green generation. a,** Pump pulses from 1 GHz Ti:sapphire comb enter a 35 mm long photonic crystal fiber to create 650–1150 nm supercontinuum (SC) pulses, which are launched into an MgO:PPLN waveguide (**inset**). Replica pump pulses are co-launched into the waveguide and overlapped with infrared supercontinuum components to achieve a broadband SHG and SFM comb, which is coupled into a Fabry-Pérot etalon to produce a 30 GHz astrocomb. **b,** Time-resolved output from the MgO:PPLN waveguide illustrating the enhanced spectral coverage and intensity achieved from the SFM process. **c,** Photograph of the UV to blue-green supercontinuum generated by the MgO:PPLN waveguide at optimal delay between the pump and supercontinuum pulses. **d,** Spectra of the fundamental pulses (red shading) and the SHG / SFM field (blue shading) generated in the MgO:PPLN waveguide. The shape and bandwidth of the SHG / SFM spectrum after Fabry-Pérot filtering (solid blue line) is preserved.

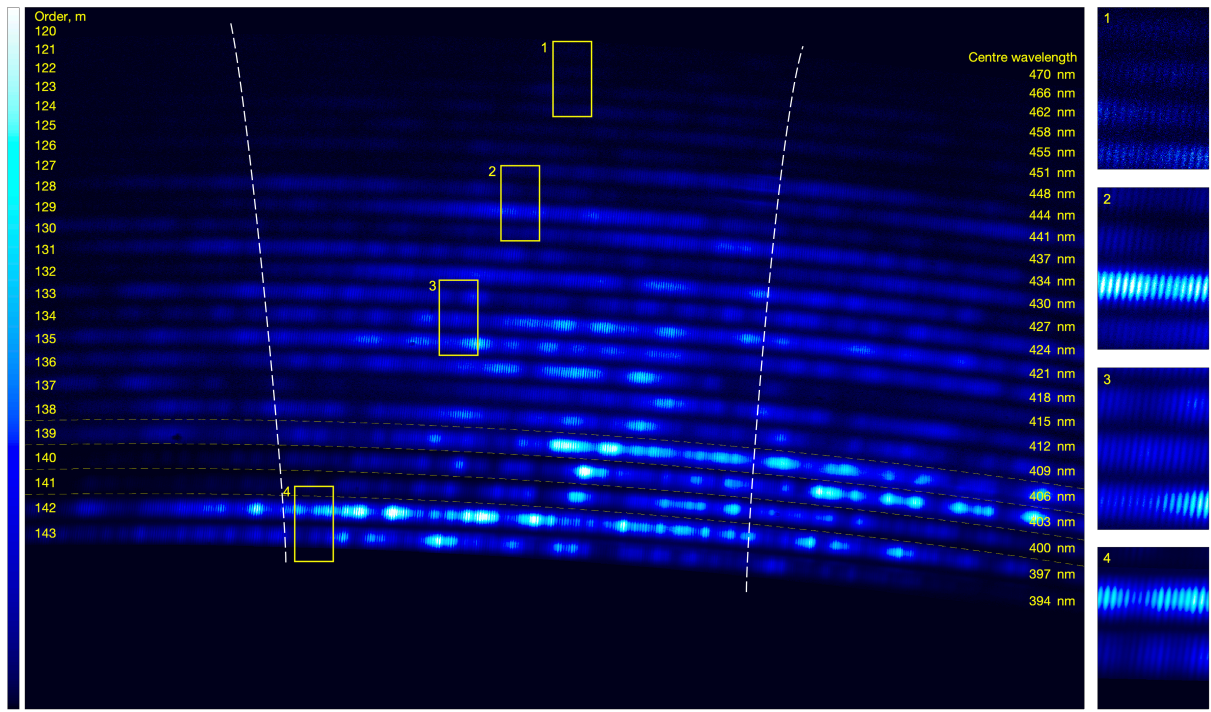

**Fig. 3. Comb-mode-resolving echellogram.** Main image, 20.2 megapixel camera frame recorded from a cross-dispersion spectrograph, with annotations indicating the index and center wavelength of each echelle order. The free spectral range of each order is 3–4 nm, equivalent to approximately 200 comb modes and indicated by the dashed white lines. Insets, magnified regions of the echellogram confirming that the 30 GHz comb modes are resolved across the full range of modes orders 120–143, corresponding to wavelengths from 392–472 nm. To compensate for the 8-bit dynamic range of the camera, orders 139–143, which are associated with the intense pump SHG light, were recorded at lower gains than orders 120–138. Regions bounded by yellow dashed lines share the same camera gain. In the main image, the color map is linear with intensity.

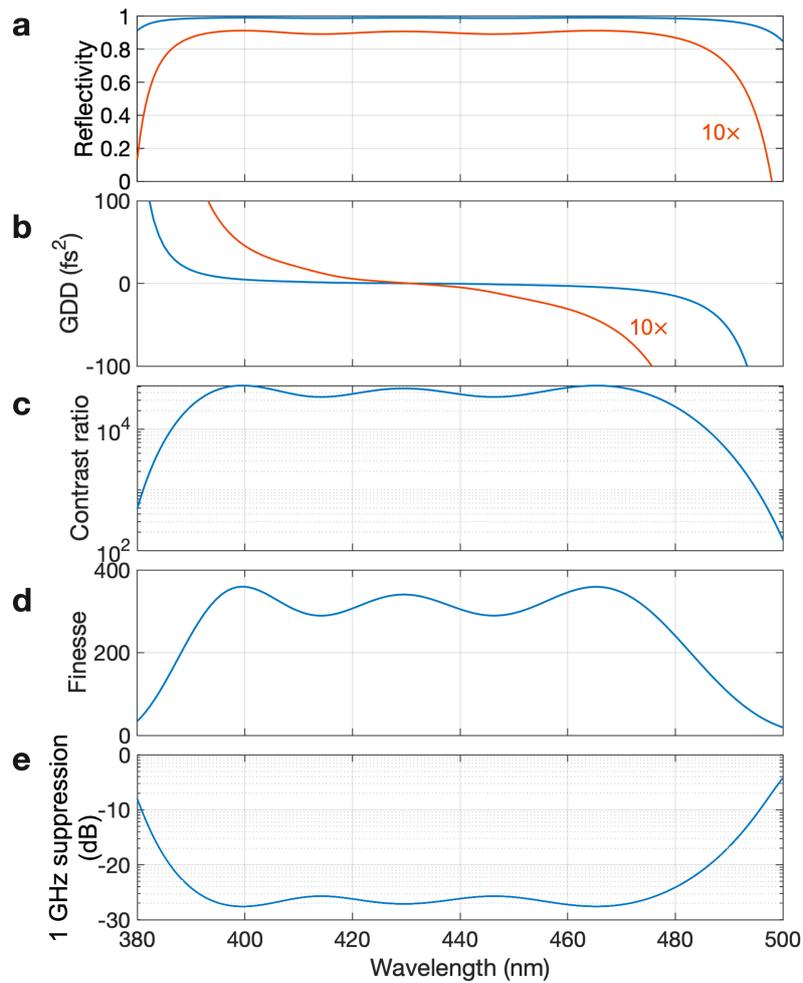

**Fig. 4. Fabry-Pérot etalon performance. a,** Wavelength dependent mirror reflectivity, with coatings having a nominal 99% reflectivity from 390–470 nm. **b,** Group-delay dispersion (GDD) of the mirror coating. **c,** Contrast ratio of the Fabry-Pérot etalon, defined as the ratio of maximum to minimum transmission. **d,** Ideal Fabry-Pérot etalon finesse. **e,** Nearest-neighbor mode suppression.